\documentclass[aps,pre,showpacs,groupedaddress]{revtex4}%PRE
\usepackage{amsfonts}
\usepackage{mathrsfs}
\usepackage{graphicx}
\usepackage{amsmath}
\usepackage{amssymb}
\usepackage{amsbsy}

\begin{document}

\title{Counting solutions from finite samplings}

\author{Haiping Huang$^{1,2}$ and Haijun Zhou$^{1}$}
%\email{hphuang@itp.ac.cn}
\affiliation{$^1$State Key Laboratory of Theoretical Physics,
Institute of Theoretical Physics, Chinese Academy of Sciences,
Beijing 100190, China\\
$^2$Department of Physics, The Hong Kong University of Science and
Technology, Hong Kong, China}
\date{\today}

\begin{abstract}
We formulate the solution counting problem within the framework of
inverse Ising problem and use fast belief propagation equations to
estimate the entropy whose value provides an estimate on the true
one. We test this idea on both diluted models (random $2$-SAT and
$3$-SAT problems) and fully-connected model (binary perceptron), and
show that when the constraint density is small, this estimate can be
very close to the true value. The information stored by the
salamander retina under the natural movie stimuli can also be
estimated and our result is consistent with that obtained by Monte
Carlo method. Of particular significance is sizes of other
metastable states for this real neuronal network are predicted.

\end{abstract}

\pacs{84.35.+i, 02.50.Tt, 75.10.Nr, 89.70.Cf}
 \maketitle

%%%%%%%%%%%%%%%%%%%%%%%%%%%%%%%%%%%%%%%%%%%%%%%%%%%%%%%%%%%%%%%%%
\section{Introduction}
Counting the number of solutions for random constraint satisfaction
problems is a very important and nontrivial problem which belongs to
\#P-complete class in computational complexity~\cite{Valiant-1979}
and is much harder than determining whether a random formula has any
solutions. In practice, we can only sample a very small part of a
huge solution space which contains an exponential number of
solutions. However, can we predict the number of solutions in the
whole solution space only based on a finite number of sampled
solutions? This issue has generated broad interests across a variety
of different disciplines such as computer science, probabilistic
reasoning, statistical physics and computational
biology~\cite{Roth-1996,Selman-2007,Kroc-2008,Favier-2009,Montanari-2009,Nature-06,Bialek-09ep,Cocco-2011,Opper-2011}.
An efficient sample-based counting strategy was proposed in
Ref.~\cite{Selman-2007}. This strategy successively sets the most
balanced variable until an exact counter is feasible on the reduced
formula, and provides a lower bound on the true count.
Alternatively, we address the solution counting problem within the
framework of inverse Ising problem in examples of diluted
models\textemdash random $K$-SAT problems and fully-connected
model\textemdash binary perceptron, and show that our method yields
an estimate whose value could be very close to the true count when
the constraint density is small. The constraint density is defined
as the ratio of the number of constraints to that of variables in
the system.

The inverse Ising problem~\cite{Ack-1985} has recently attracted
much attention not only in the development of fast mean field
inverse algorithms~\cite{Mezard-09,Roudi-2009,SM-09,Cocco-2011} but
also in modeling vast amounts of biological
data~\cite{Nature-06,Weigt-2009,Cocco-09,Mora-2010}. The pairwise
Ising model is able to capture most of the correlation structure of
the real neuronal network activity and is much more informative than
the independent model where each neuron is assumed to fire
independently~\cite{Nature-06,Tang-2008}. The observed collective
behavior of a large neuronal network results from interactions of
many individual neurons. The joint activity patterns for a retina
under naturalistic stimuli were reported to convey information about
the visual stimuli~\cite{Tkacik-2010,Bialek-09ep}. Estimating the
information stored by the real neuronal network directly from data
remains an open and important issue. We show in this work the
information can be estimated reliably and sizes of metastable states
for the neuronal network can also be predicted.

The paper is organized as follows. The inverse Ising problem is
introduced in Sec.~\ref{InvIsing}, together with a brief description
of the susceptibility propagation algorithm used to infer the
disordered Ising model. In Sec.~\ref{Method}, we present the belief
propagation to estimate the entropy from the data and apply this
method to predict the entropies of four different examples only from
a limited number of samplings. Finally, the conclusion suggests some
implications of our study as well as potential applications of the
presented methodology.

%%%%%%%%%%%%%%%%%%%%%%%%%%%%%%%%%%%%%%%%%%%%%%%%%%%%%%%%%%%%%%%%%%%
\section{the inverse Ising problem}
\label{InvIsing}
 For a system of $N$ variables, one can collect $P$ configurations or solutions
$\{\sigma_{i}^{\nu}\}(i=1,\ldots,N;\nu=1,\ldots,P)$ either from real
biological experiments (e.g., spike trains in multi-electrode array
recordings~\cite{Nature-06}) or from random walks in the solution
space of a model. We assume $\sigma_{i}$ takes Ising-type value
$\pm1$. The task of the inverse Ising problem is to find couplings
$\{J_{ij}\}$ and fields $\{h_{i}\}$ to construct a minimal model
\begin{equation}\label{Ising}
    P_{{\rm
Ising}}(\boldsymbol\sigma)=\frac{1}{Z}\exp\left[\sum_{i<j}J_{ij}\sigma_{i}\sigma_{j}+\sum_{i}h_{i}\sigma_{i}\right]
\end{equation}
such that its magnetizations and pairwise correlations are
compatible with those measured, i.e., $\left<\sigma_{i}\right>_{{\rm
Ising}}=\left<\sigma_{i}\right>_{{\rm data}},
\left<\sigma_{i}\sigma_{j}\right>_{{\rm
Ising}}=\left<\sigma_{i}\sigma_{j}\right>_{{\rm data}}$. $Z$ is the
partition function and the inverse temperature $\beta=1$ as it can
be absorbed in the strength of couplings and fields. Hereafter, we
define the measured magnetization and connected correlation as
$m_{i}\equiv\left<\sigma_{i}\right>_{{\rm data}}$ and
$C_{ij}\equiv\left<\sigma_{i}\sigma_{j}\right>_{{\rm
data}}-m_{i}m_{j}$ respectively where $\left<\cdots\right>_{{\rm
data}}$ denotes the average over the sampled configurations or
solutions.

We use susceptibility propagation (SusProp) to infer the couplings
and fields. SusProp passes messages along the oriented edges of the
network by iterative updating. To run SusProp, two kinds of messages
are needed. One is the cavity magnetization of variable $i$ in the
absence of variable $j$ denoted as $m_{i\rightarrow j}$; the other
is the cavity susceptibility $g_{i\rightarrow j,k}$ that is the
response of cavity field of variable $i$ without variable $j$ to a
local perturbation of external field of variable
$k$~\cite{Mezard-09}. The update rule can be derived using belief
propagation Eq.~(\ref{bpIsing}) and fluctuation-response
relation~\cite{Mezard-09,Marinari-2010,Huang-2010b} and reads as
follows~\cite{Huang-2010b}:
%\begin{widetext}
\begin{subequations}\label{SusP}
\begin{align}
m_{i\rightarrow j}&=\frac{m_{i}-m_{j\rightarrow i}\tanh J_{ij}}{1-m_{i}m_{j\rightarrow i}\tanh J_{ij}}\\
g_{i\rightarrow j,k}&=\delta_{ik}+\sum_{l\in \partial i\backslash
j}\frac{1-m_{l\rightarrow i}^{2}}{1-(m_{l\rightarrow i}\tanh
J_{li})^{2}}
\tanh J_{li}g_{l\rightarrow i,k}\\
J_{ij}^{{\rm
new}}&=\frac{\epsilon}{2}\log\left(\frac{(1+\tilde{C_{ij}})(1-m_{i\rightarrow
j}m_{j\rightarrow i})}
{(1-\tilde{C_{ij}})(1+m_{i\rightarrow j}m_{j\rightarrow i})}\right)+(1-\epsilon)J_{ij}^{{\rm old}}\\
\tilde{C_{ij}}&=\frac{C_{ij}-(1-m_{i}^{2})g_{i\rightarrow
j,j}}{g_{j\rightarrow i,j}}+m_{i}m_{j}
\end{align}
\end{subequations}
%\end{widetext}
where $\partial i\backslash j$ denotes neighbors of variable $i$
except $j$, $\delta_{ik}$ is the Kronecker delta function and
$\epsilon\in[0,1]$ is introduced as a damping factor and should be
appropriately chosen to prevent the absolute updated $\tanh(J_{ij})$
from being larger than $1$. In practice, all couplings are initially
set to be zero and for every directed edge of the network, the
message $m_{i\rightarrow j}$ is randomly initialized in the interval
$[-1.0,1.0]$ and $g_{i\rightarrow j,k}=0$ if $i\neq k$ and $1.0$
otherwise. The SusProp rule Eq.~(\ref{SusP}) is then iterated until
either the inferred couplings converge within a predefined precision
$\eta^{(1)}$ or the preset maximal number of iterations $\mathcal
{T}_{{\rm max}}^{(1)}$ is exceeded. After the set of couplings is
obtained, the fields are inferred via $
h_{i}=\tanh^{-1}(m_{i})-\sum_{l\in\partial i}\tanh^{-1}\left[\tanh
J_{li}m_{l\rightarrow i}\right]$.

To ensure a reliable estimate of the parameters, we define a
convergence fraction $R$ as the ratio of the number of converged
couplings to the total number of edges in the network. In the
non-convergent case, we take the inferred parameters corresponding
to $R_{\rm max}=\max\{R_{t},t=1,\ldots,\mathcal {T}_{{\rm
max}}^{(1)}\}$ where $R_{t}$ is the convergence fraction of $t$-th
iteration. $R_{{\rm max}}=1.0$ if the update rule converges. For an
inverse problem, $\{m_{i},C_{ij}\}$ serve as inputs to the update
rule, and they are computed from $P$ sampled solutions or
configurations. We use stochastic local search algorithms to sample
the solution space of random $K$-SAT ($K=2,3$ here) formulas and
that of the binary perceptron. For the retinal network, the
configurations were obtained from the spike trains in the
multi-electrode recording experiments (data courtesy of Gasper
Tkacik, Refs.~\cite{Nature-06,Bialek-09ep}).

\section{Estimating the entropy from the data}
\label{Method}
 We derive the entropy of the constructed Ising model
Eq.~(\ref{Ising}) under Bethe approximation (also called cavity
method~\cite{cavity-2001}) assuming sufficiently weak interactions
among variables. We compute the entropy through site contributions
$\Delta S_{i}$ and edge contributions $\Delta S_{ij}$ as $S_{{\rm
Ising}}=Ns_{{\rm Ising}}=\sum_{i}\Delta
S_{i}-\sum_{\left<ij\right>}\Delta S_{\left<ij\right>}$:
\begin{widetext}
\begin{subequations}\label{entropy}
\begin{align}
\begin{split}
    \Delta S_{i}&=\log Z_{i}-\frac{1}{Z_{i}}\Biggl[h_{i}e^{h_{i}}\prod_{l\in\partial i}\cosh J_{li}(1+\tanh J_{li}m_{l\rightarrow
    i})-h_{i}e^{-h_{i}}\prod_{l\in\partial i}\cosh J_{li}(1-\tanh J_{li}m_{l\rightarrow i})\\
    &+e^{h_{i}}\sum_{l\in\partial i}\Bigl[J_{li}\sinh J_{li}(1+\tanh J_{li}m_{l\rightarrow i})+J_{li}\cosh J_{li}(1-\tanh^{2}J_{li})m_{l\rightarrow i}\Bigr]
    \cdot\prod_{j\in\partial i\backslash l}\cosh J_{ij}(1+\tanh J_{ij}m_{j\rightarrow i})\\
    &+e^{-h_{i}}\sum_{l\in\partial i}\Bigl[J_{li}\sinh J_{li}(1-\tanh J_{li}m_{l\rightarrow i})-J_{li}\cosh J_{li}(1-\tanh^{2}J_{li})m_{l\rightarrow i}\Bigr]
    \cdot\prod_{j\in\partial i\backslash l}\cosh J_{ij}(1-\tanh
J_{ij}m_{j\rightarrow i})\Biggr]
    \end{split}\\
    \Delta S_{\left<ij\right>}&=\log Z_{ij}-J_{ij}\frac{\tanh J_{ij}+m_{i\rightarrow j}m_{j\rightarrow i}}{1+\tanh J_{ij}m_{i\rightarrow j}m_{j\rightarrow i}}
\end{align}
\end{subequations}
\end{widetext}
where $\partial i\backslash l$ denotes neighbors of variable $i$
except $l$. $Z_{i}=e^{h_{i}}\prod_{l\in\partial i}\cosh
J_{li}(1+\hat{m}_{l\rightarrow i})+e^{-h_{i}}\prod_{l\in\partial
i}\cosh J_{li}(1-\hat{m}_{l\rightarrow i})$ and $Z_{ij}=\cosh
J_{ij}\left(1+\tanh J_{ij}m_{i\rightarrow j}m_{j\rightarrow
i}\right)$. The cavity magnetization $m_{i\rightarrow j}$ obeys
simple recursive equations:
\begin{subequations}\label{bpIsing}
\begin{align}
    m_{i\rightarrow j}&=\frac{e^{h_{i}}\prod_{l\in\partial i\backslash j}(1+\hat{m}_{l\rightarrow i})-e^{-h_{i}}\prod_{l\in\partial i\backslash j}(1-\hat{m}_{l\rightarrow i})}
    {e^{h_{i}}\prod_{l\in\partial i\backslash j}(1+\hat{m}_{l\rightarrow i})+e^{-h_{i}}\prod_{l\in\partial i\backslash j}(1-\hat{m}_{l\rightarrow i})}\\
    \hat{m}_{l\rightarrow i}&=\tanh J_{li}m_{l\rightarrow i}
\end{align}
\end{subequations}
We first randomly initialize $m_{i\rightarrow j}\in[-1.0,1.0]$ for
every directed edge of the reconstructed network, then iterate
Eq.~(\ref{bpIsing}) until all messages converge within the precision
$\eta^{(2)}$ or the maximal number of iterations $\mathcal {T}_{{\rm
max}}^{(2)}$ is reached. From the fixed point, the entropy can be
computed via Eq.~(\ref{entropy}). The case where some variable, say
$i$ is positively frozen, i.e., corresponding measured magnetization
$m_{i}=1.0$, can also be handled. In this case, $h_{i}=+\infty$, and
$\Delta S_{i}$ is reduced to be $\log
Z_{i}^{'}-\frac{Z_{i}^{''}}{Z_{i}^{'}}$ where
$Z_{i}^{'}=\prod_{l\in\partial i}\cosh J_{li}\left(1+\tanh
J_{li}m_{l\rightarrow i}\right)$ and $Z_{i}^{''}=\sum_{l\in\partial
i}\Bigl[J_{li}\sinh J_{li}(1+\tanh J_{li}m_{l\rightarrow
i})+J_{li}\cosh J_{li}(1-\tanh^{2}J_{li})m_{l\rightarrow
i}\Bigr]\cdot\prod_{j\in\partial i\backslash l}\cosh J_{ij}(1+\tanh
J_{ij}m_{j\rightarrow i})$. The edge contribution remains unchanged.
The negatively frozen case is similarly treated. In numerical
simulations, we adopt $\mathcal {T}_{{\rm max}}^{(2)}=500$,
$\eta^{(2)}=10^{-4}$, $\mathcal {T}_{{\rm max}}^{(1)}=2000$.
$\eta^{(1)}$ as well as $\epsilon$ depends on the following specific
applications.

We remark here that Eq.~(\ref{entropy}) is used specifically for the
solution counting problem where we now have known the magnetizations
and correlations and additionally some frozen cases (some $m_{i}=+1$
or $-1$) should be treated. On the other hand, the coupling or field
distributions depend on the collected data and Eq.~(\ref{entropy})
is derived only under the weakly-coupled approximation but the fully
connected topology is reserved. The first point is, the entropy we
try to estimate is not only for two-body interaction system (e.g.,
random $2$-SAT) but also for three-body interaction system and
densely-interacted system (e.g., the binary perceptron where each
constraint involves all variables of the system). The second point
is, the sampled solutions come from the zero energy ground state and
the sampling process is always confined in a single cluster
(solutions in it are connected with each other by single variable
flips).

All underlying parameters of pairwise Ising model are predicted
directly from the observed data and the entropy of the original
model is estimated based on the constructed Ising model. We
emphasize here that two layers of approximations are made. The first
one is the disordered Ising model Eq.~(\ref{Ising}) is used to
approximate the original model. When estimating the entropy from the
data, we actually do not know the original model. The second layer
is we use mean-field methods, specifically the message passing
algorithms to infer the underlying parameters of the pairwise Ising
model. Since the computational complexity of SusProp is $\mathcal
{O}(N^{3})$ for the fully-connected network, we focus on small size
networks with $N$ of order $\mathcal {O}(10^{2})$. When the
constraint density is small, the efficiency of our methodology is
supported by two concrete examples: random $K$-SAT problem and the
binary perceptron. For these two examples, we use $s_{{\rm true}}$
to represent the entropy density computed by belief propagation with
the knowledge of the original model (for details, see
Ref.~\cite{Zhou-2009} for random $K$-SAT problem and
Ref.~\cite{Zecchina-2006} for binary perceptron). To show the
efficiency of the pairwise Ising model, we also compute the
independent entropy $S_{{\rm ind}}=Ns_{{\rm
ind}}=-\sum_{i}\left[\frac{1+m_{i}}{2}\log\frac{1+m_{i}}{2}+\frac{1-m_{i}}{2}\log\frac{1-m_{i}}{2}\right]$
assuming $P(\boldsymbol\sigma)\approx\prod_{i}P_{i}(\sigma_{i})$.
For retinal network, we could neither know the true model underlying
the network nor get the true value for the entropy (when the network
is large). Therefore we just compare the result obtained by our
current fast belief propagation with that obtained by time-consuming
Monte Carlo method and show that the belief propagation not only
reproduces the entropy value evaluated by Monte Carlo method but
also yields rich information about the metastable states which are
relevant for neuronal population
coding~\cite{Bialek-09ep,Tkacik-2010}. In this case, we denote
$s_{{\rm BP}}$ as the entropy density estimated by belief
propagation and $s_{{\rm MC}}$ estimated by Monte Carlo method. Note
that both belief propagation and Monte Carlo method under the
reconstructed Ising model yield approximate value for the true
entropy since we consider only up to second-order correlations in
the observed data while the system may develop higher-order
correlations in its solution space or energy landscape. The
different natures of these examples imply wide applications of our
methodology to evaluate the entropy of an unknown model with only a
limited number of samplings.
%%%%%%%%%%%%%%%%%%%%%%%%%%%%%%%%%%%%%%%%%%%%%%%%%%%%%%%%%%%%%%%%%%%%

\begin{figure}
          \includegraphics[bb=9 14 311 220,scale=0.85]{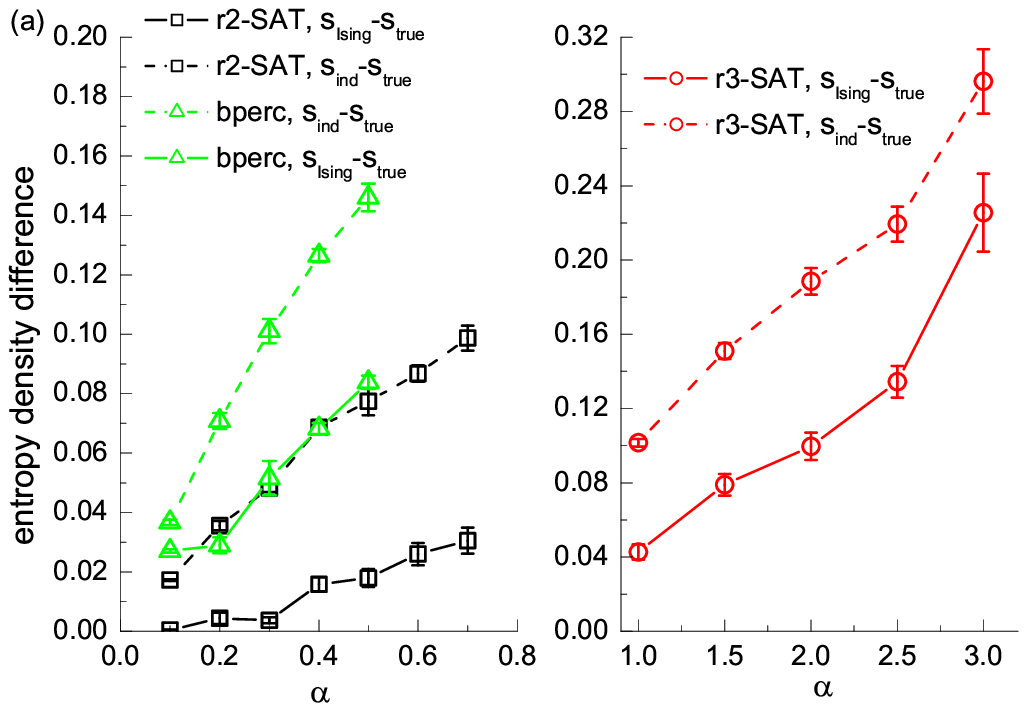}
     \hskip .2cm
     \includegraphics[bb=14 15 252 145,scale=1.0]{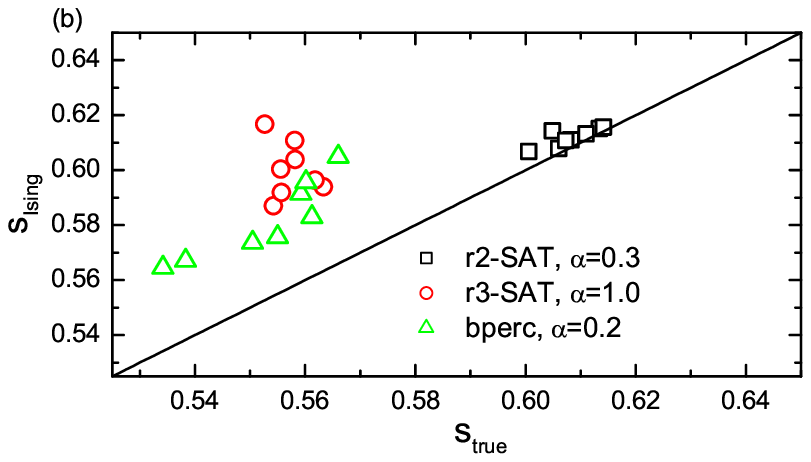}\vskip .2cm
  \caption{(Color online)
     The quality of the pairwise and independent models versus constraint density $\alpha$. (a) Entropy density difference versus $\alpha$.
     The data points connected by dashed line are the differences between $s_{{\rm ind}}$
     and $s_{{\rm true}}$, while those connected by solid line are the differences $s_{{\rm Ising}}-s_{{\rm true}}$. The number of variables $N=100$ for random
     $K$-SAT (r$2$-SAT or r$3$-SAT)
     problem and $101$ for binary perceptron (bperc). $s_{{\rm true}}$ is
     computed with the knowledge of the original model using belief
     propagation~\cite{Zhou-2009,Zecchina-2006}.
     Each point represents the average over eight
     random samples. (b) Scatter
     plot comparing $s_{{\rm Ising}}$ with $s_{{\rm true}}$. The full line indicates equality.
   }\label{compa}
 \end{figure}
\subsection{Random $K$-SAT problem}
\label{sat}
 The random $K$-SAT problem is finding a solution (an assignment of
$N$ boolean variables) satisfying a random formula composed of
logical AND of $M$ constraints~\cite{Monasson-1996}. Each constraint
is a logical OR function of $K$ randomly chosen distinct variables
(either directed or negated with equal probability). The constraint
density $\alpha=M/N$. For $K=2$, the threshold separating a SAT
phase from an UNSAT phase was confirmed to be $\alpha_{s}=1$ below
which the solution space is ergodic and a simple local search
algorithm can easily identify a solution~\cite{Monasson-1996}. For
$K=3$, the estimated threshold $\alpha_{s}\simeq4.267$ below which
the solution space exhibits richer
structures~\cite{Krzakala-PNAS-2007}. The dynamical transition point
locates at $\alpha_{d}\simeq3.86$ and separates the ergodic phase
from non-ergodic phase. We use {\tt SEQSAT} algorithm of
Ref.~\cite{Zhou-2010epjb} to first find a solution for a given
$\alpha$, then $10^{8}N$ single variable flips are performed in the
current solution space, after that we perform random walks in the
current solution
 space to sample one solution every $10^4$ steps. Each step
 involves $N$ attempts to move from one solution to its adjacent one
 by single variable flip, i.e., a randomly chosen variable is
 flipped and if the new configuration is a solution, the flip is
 accepted with probability $1/2$; otherwise the movement is
 rejected. We sample totally $P=10^{5}$ solutions to estimate the entropy. We choose
 $\epsilon=0.128,\eta^{(1)}=10^{-3}$ for random $2$-SAT and $\epsilon=0.002,\eta^{(1)}=10^{-4}$ for random
 $3$-SAT. Results are reported in Fig.~\ref{compa}. When $\alpha$ is small, our method can predict the true entropy very well
 especially for $K=2$ which can be actually transformed into a pairwise Ising model. As $\alpha$ increases, the difference between $s_{{\rm
 Ising}}$ and $s_{{\rm
 true}}$~\cite{Zhou-2009} becomes large and this deviation is more obvious for $K=3$,
 which manifests the presence of higher-order
 correlations in the solution space. At high
 $\alpha$ (e.g., $\alpha=3.0$ for $K=3$), the belief propagation Eq.~(\ref{bpIsing}) would yield multiple fixed
 points. This signals ergodicity breaking phenomenon in the
 energy landscape of the constructed Ising model or indicates that long range correlations develop in the original system~\cite{Montanari-2009}, although our samplings are still confined in a
 single cluster of the original model, as a result, the predicted entropy becomes rather inaccurate compared with the true one computed under the original model.

\subsection{Binary perceptron}
\label{bperc}
The binary perceptron with $N$ binary weights
connecting $N$ input nodes to a single output node performs a random
classification of $\alpha N$ random binary patterns
$\{\xi_{i}^{\mu}\}(i=1,\ldots,N;\mu=1,\ldots,\alpha N)$. The
critical constraint density $\alpha_{s}\simeq0.83$ below which the
solution space is non-empty~\cite{Krauth-1989}. Given an input
pattern $\boldsymbol{\xi}^{\mu}$, if the actual output $o^{\mu}={\rm
sgn}\left(\sum_{i=1}^{N}\sigma_{i}\xi_{i}^{\mu}\right)$ is equal to
the desired output $o_{0}^{\mu}$ assigned a value $\pm 1$ with equal
probabilities, the configuration $\boldsymbol{\sigma}$ learns this
pattern. The solution space of the binary perceptron consists of all
configurations learning $\alpha N$ random patterns. Before sampling,
we first learn $\alpha N$ patterns using {\tt DWF}
 algorithm of Ref.~\cite{Huang-2010jstat}. The sampling procedure is
 the same as that used for random $K$-SAT problems. In numerical
 simulations, we choose $\epsilon=0.001,\eta^{(1)}=10^{-4}$. The
 deviation of estimated $s_{{\rm Ising}}$ from
 $s_{{\rm true}}$~\cite{Zecchina-2006} is plotted against $\alpha$. For
 small $\alpha$, our method can predict the true entropy well
 without the knowledge of the original model. The large deviation shown in Fig.~\ref{compa}
 at high $\alpha$ implies higher-order correlations start to dominate the solution space.

\begin{figure}
          \includegraphics[bb=11 13 245 137,scale=1.0]{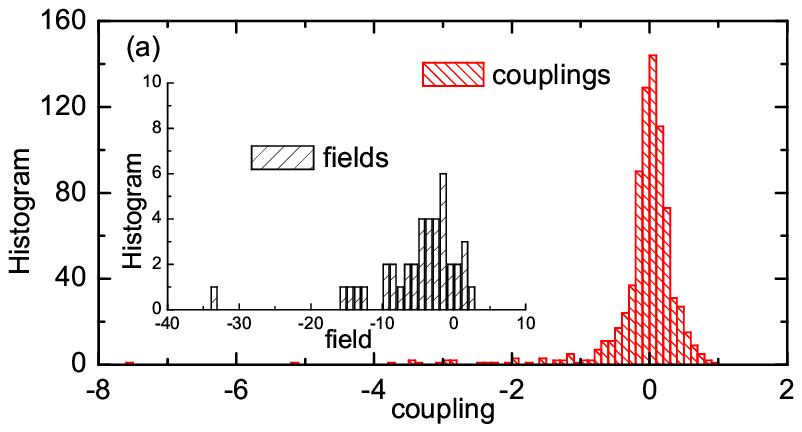}
          \hskip .1cm
     \includegraphics[bb=11 16 282 216,scale=0.8]{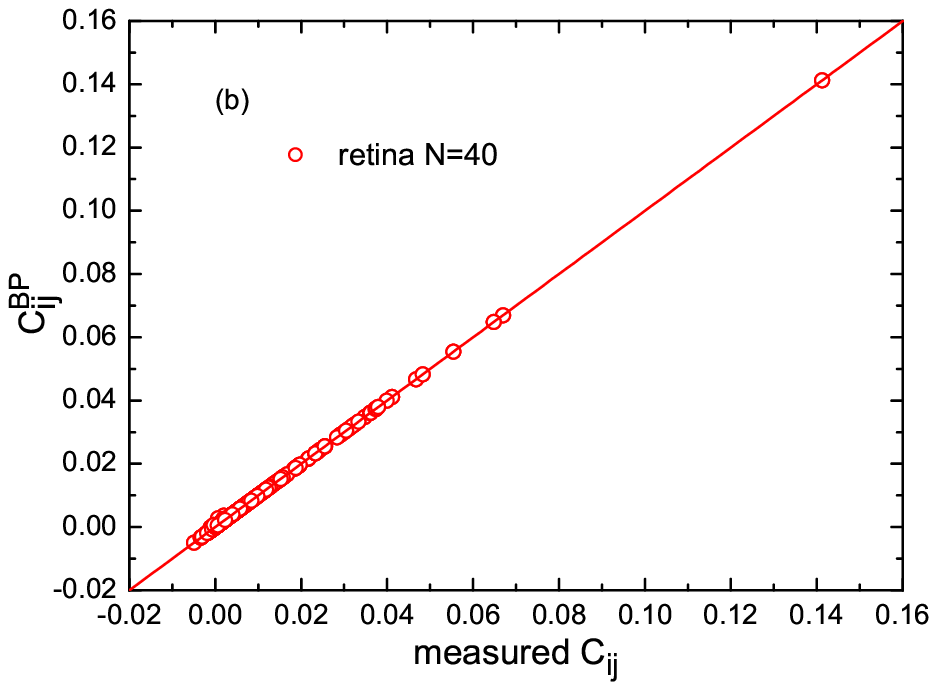}\vskip .2cm
  \caption{(Color online)
     (a) Histograms of inferred couplings and fields (inset) for the
     retinal network with the number of neurons $N=40$ (data
     courtesy of Gasper Tkacik, Refs.~\cite{Nature-06,Bialek-09ep}).
     To infer the network, we choose
     $\epsilon=0.128,\eta^{(1)}=10^{-3}$. The network is inferred at
     $R_{{\rm max}}=0.959$. (b) Reconstructed $C_{ij}^{{\rm BP}}$ using belief propagation Eq.~(\ref{Corre}) versus the measured one.
     We only show the case $i\neq j$ since $C_{ii}^{{\rm BP}}=1-(m_{i}^{{\rm
     BP}})^{2}$. The full line indicates equality.
   }\label{histo}
   \end{figure}

   \begin{table}
\caption{Estimated entropy density $s_{{\rm BP}}$ for the inferred
retinal network through belief propagation Eq.~(\ref{bpIsing}).
$q_{0}=\frac{1}{N}\sum_{i}m_{i}^{{\rm BP}}m_{i}$ where $m_{i}^{{\rm
BP}}$ is computed from the fixed point of belief propagation
Eq.~(\ref{bpIsing}). The self-overlap
$q_{1}=\frac{1}{N}\sum_{i}m_{i}^{2}\simeq0.906828$. The last column
gives the probability of appearance for each fixed point during
$1000$ runs of belief propagation with the same inferred
parameters.}\label{table:retina}
    \begin{tabular}{c  c  c  c}
    \hline
    \hline
    $s_{{\rm BP}}$ &   $q_{0}$ & prob. app  \\
     \hline
    0.09771     & 0.906816      & 0.089\\ 0.1093    & 0.512411      & 0.075\\
0.1224      &0.313566       & 0.066\\ 0.1634        & 0.281124      & 0.400\\ 0.1765        & 0.021960      & 0.354\\
0.1918      & 0.264979      & 0.016\\
    \hline
    \hline
    \end{tabular}
    \end{table}

\subsection{Retinal network}
\label{retina} A recording of the activity of $40$ neurons in a
salamander retina under natural movie stimuli could also be analyzed
within the current setting. The total effective number of samplings
$P\simeq7\times10^{4}$~\cite{Bialek-09ep}. Our estimated entropy
density for the retinal network is $s_{{\rm BP}}\simeq0.09771$
consistent with that obtained by Gasper Tkacik
et.al~\cite{Bialek-09ep} using Monte Carlo method which produces an
estimate $s_{{\rm MC}}\simeq0.09479$ but is rather time consuming
for large $N$. Our result implies that the retina under the
naturalistic movie stimuli stores $e^{Ns_{{\rm BP}}}\simeq50$
effective configurations. If we approximate the true entropy using
that calculated by belief propagation (Eq.~(\ref{entropy}) and
Eq.~(\ref{bpIsing})) or Monte Carlo method, then the
multi-information measuring the total amount of correlations in the
network~\cite{Bialek-03} $I_{{\rm BP}}=s_{{\rm ind}}-s_{{\rm BP}}$
or $I_{{\rm MC}}=s_{{\rm ind}}-s_{{\rm MC}}$ where $s_{{\rm ind}}$
is the independent entropy density. The result is that the
difference between these two multi-information values $I_{{\rm
MC}}-I_{{\rm BP}}\simeq0.00292$. The histogram of inferred
parameters is shown in Fig.~\ref{histo} (a). Note that most of
predicted couplings concentrate around zero value with a long tail
of distribution for large negative couplings whose weights are
rather small. Most of the predicted fields are negative since most
of the neurons are silent across the movie presentations. Using the
same inferred parameters, we run belief propagation
Eq.~(\ref{bpIsing}) $1000$ times from different random
initializations. Several fixed points are found and one of them is
consistent with the previous result~\cite{Bialek-09ep} (see
Table~\ref{table:retina}). These fixed points represent different
metastable states and the entropy measures the capacity of neurons
to convey information about the visual stimulus which contains
high-order correlation structure. The visual information could be
encoded by identity of the basin of attraction~\cite{Tkacik-2010}
and these predicted metastable states may code for specific stimulus
features. Therefore, the information of the inputs to the retina can
be stored in the couplings and fields which generate a free energy
landscape with multiple metastable states for redundant error
correction~\cite{Tkacik-2010}. Future research on neuronal
population coding needs to elucidate this point.

In Fig.~\ref{histo} (b), we verify that the inferred pairwise Ising
model reproduces the measured connected correlations with very good
agreement. The reconstructed correlations $\{C_{ij}^{{\rm BP}}\}$
can be computed by the following message passing
algorithm~\cite{Higuchi-2010,Huang-2010b}:
\begin{subequations}\label{bpC}
\begin{align}
    m_{i\rightarrow j}&=\frac{e^{h_{i}}\prod_{l\in\partial i\backslash j}(1+\hat{m}_{l\rightarrow i})-e^{-h_{i}}\prod_{l\in\partial i\backslash j}(1-\hat{m}_{l\rightarrow i})}
    {e^{h_{i}}\prod_{l\in\partial i\backslash j}(1+\hat{m}_{l\rightarrow i})+e^{-h_{i}}\prod_{l\in\partial i\backslash j}(1-\hat{m}_{l\rightarrow i})}\\
    \hat{m}_{l\rightarrow i}&=\tanh J_{li}m_{l\rightarrow i}\\
    g_{i\rightarrow j,k}&=\delta_{ik}+\sum_{l\in \partial i\backslash
j}\frac{1-m_{l\rightarrow i}^{2}}{1-(m_{l\rightarrow i}\tanh
J_{li})^{2}} \tanh J_{li}g_{l\rightarrow i,k}
\end{align}
\end{subequations}
where two kinds of messages, $m_{i\rightarrow j}$ and
$g_{i\rightarrow j,k}$ are updated. Once both messages for each
directed link in the network are converged, i.e., iteration of
Eq.~(\ref{bpC}) reaches fixed point, we compute the predicted
connected correlations $\{C_{ij}^{{\rm BP}}\}$ via
\begin{subequations}\label{Corre}
\begin{align}
    C_{ij}^{{\rm BP}}&=(\tilde{C_{ij}}-m_{i}m_{j})g_{j\rightarrow i,j}+(1-m_{i}^{2})g_{i\rightarrow j,j}\\
    \tilde{C_{ij}}&=\frac{\tanh J_{ij}+m_{i\rightarrow j}m_{j\rightarrow i}}{1+\tanh J_{ij}m_{i\rightarrow j}m_{j\rightarrow i}}
\end{align}
\end{subequations}
where the Ising model is known and all messages needed to compute
$C_{ij}^{{\rm BP}}$ including $m_{i}$ and $m_{j}$ are read from the
fixed point. This message passing strategy to evaluate the
correlations is very fast and takes tens of iterations to converge.
Remarkably, the estimated magnetizations and correlations fit those
measured very well. However, using Monte Carlo samplings to
reconstruct the correlations, we failed to reproduce those measured.
For example, after sufficient thermalization, the configuration is
sampled every $10^{4}$ Monte Carlo sweeps, then the correlations are
computed with $10^{4}$ sampled configurations. The obtained root
mean square error is about $0.23$. Possible reason is, the
reconstructed Ising model by SusProp already develops multiple
states (different fixed points of belief propagation
Eq.~(\ref{bpIsing})), therefore, at temperature $T=1.0$ and $N=40$,
as observed in our simulations, Monte Carlo samplings can have
transitions between different states yielding the average energy
density of sampled configurations $\bar{e}\simeq-3.7119$ with
fluctuation of order $0.0724$ while the state reproducing the
measured correlations in Fig.~\ref{histo} (b) has typical energy
density $-3.62954$ but smallest entropy density $s_{{\rm
BP}}\simeq0.09771$ of all observed states. Actually, in this case,
the correlation computed by Monte Carlo samplings corresponds to the
average over different states while the measured data comes from one
state of the reconstructed Ising model.

\subsection{Convergence patterns and entropy density difference versus $P$}
\label{edp}
\begin{figure}
          \includegraphics[bb=14 15 273 206,scale=0.8]{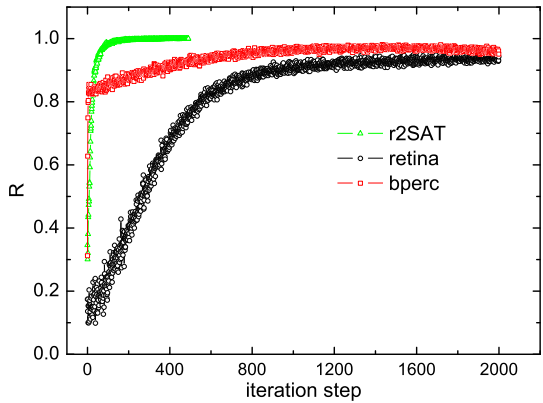}
  \caption{(Color online)
     Convergence patterns possibly appearing in the iteration of
     SusProp update rules.
   }\label{CovP}
   \end{figure}

%%%%%%%%%%%%%%%%%%%%%%%%%%%%%%%%%%%%%%%%%%%%%%%%%%%%%%%%%%%%%%%%
There exist three kinds of convergence patterns for different
iterations of SusProp rules. One is the convergence case shown by an
example of random $2$-SAT with $\alpha=0.7$; the second type is the
convergence fraction $R$ first increases then decreases (see in
Fig.~\ref{CovP} an example of binary perceptron with $\alpha=0.5$);
the last type is $R$ reaches a plateau with small fluctuations shown
by an example of retina.
%%%%%%%%%%%%%%%%%%%%%%%%%%%%%%%%%%%%%%%%%%%%%%%%%%%%%%%%%%%%%%%
\begin{figure}
          \includegraphics[bb=17 16 283 212,scale=0.85]{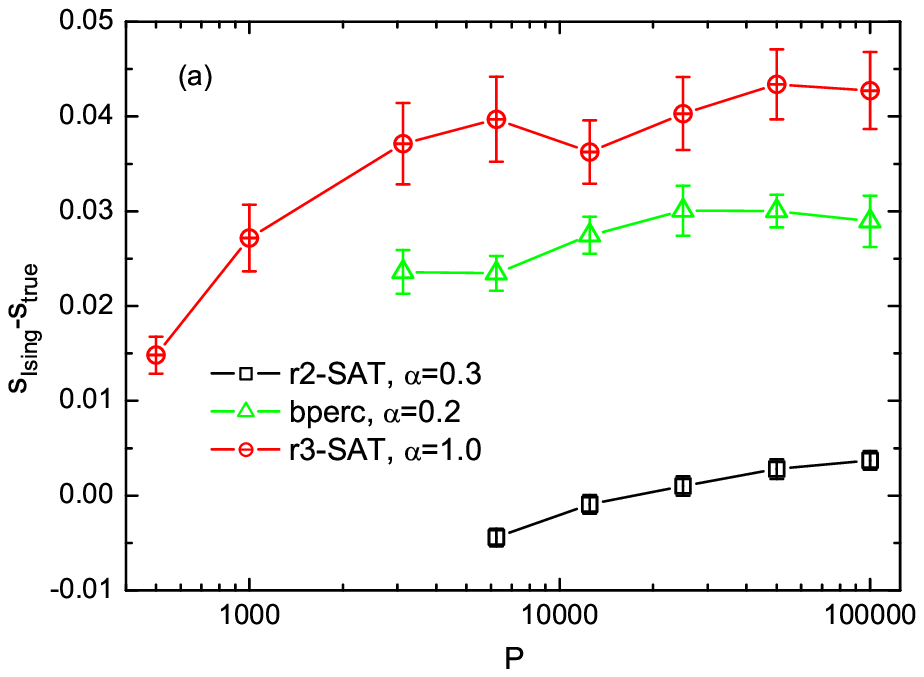}
     \hskip .2cm
     \includegraphics[bb=14 15 300 214,scale=0.85]{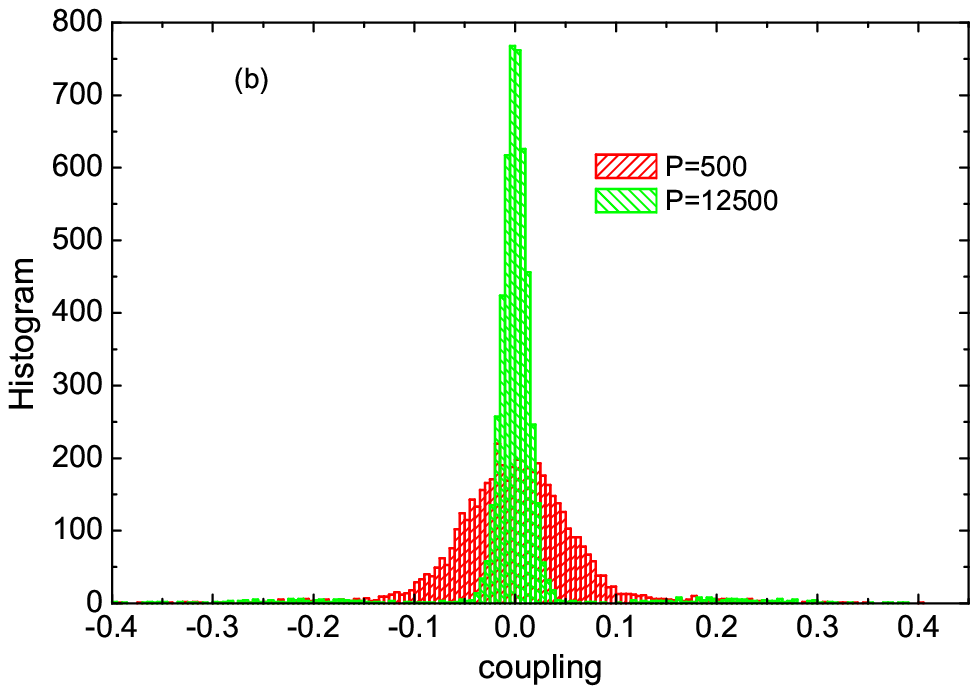}\vskip .2cm
  \caption{(Color online)
     (a) Entropy density difference versus $P$. The error bar shows the fluctuation across eight random samples. (b) The distribution
     of inferred couplings for r$3$-SAT with $\alpha=1.0$ and
     $N=100$. Two results for $P=500$ and $P=12500$ are shown.
   }\label{Edf}
 \end{figure}

 In Fig.~\ref{Edf}(a), the influence of the number of samplings $P$ on
 the estimation of entropy from the finite samplings is shown. It
 seems that the entropy density difference decreases as $P$ becomes
 small. Note that we construct the pairwise Ising model based on the
 samplings from the ground state (zero energy for these three
 constraint satisfaction problems) and the underlying graphical
 model (e.g., r$3$-SAT or bperc) may not be a pairwise Ising model.
 Our samplings are always confined in a single solution cluster and
 the quality may depend on the fine structure of the solution space~\cite{Zhou-2009pre,Hartmann-2010,Huang-2011epl}.
 This case is different from studies on the reconstruction of
 Sherrington-Kirkpatrick model at high
 temperatures~\cite{Marinari-2010}. In our current setting, when the
 sampling number $P$ decreases, the collected data seems to have
 more correlations (this may be induced by the statistical errors) and the SusProp turns out to be not converged any more (e.g., in the case of r$2$-SAT with $\alpha=0.3$), which
 can be justified from the Fig.~\ref{Edf}(b) that the inferred
 coupling distribution becomes broader with decreasing $P$ and thus the estimated entropy should take smaller values.
 As observed in Fig.~\ref{Edf}(b), once $P$ decreases down to some value, the estimated entropy value would underestimate the true one computed with the knowledge of
 the original model. On the other hand, given small $P$, the computed magnetizations and correlations
would have large statistical errors and may not correctly reflect
the correlations in the sampled solution cluster. Safely, we
 select $P=10^{5}$ to reduce the statistical error for calculating
 the magnetizations and correlations.
\section{Conclusion}
\label{Conc}
 In this work, we address the important problem of
counting the total number of solutions (configurations) based on a
limited number of sampled solutions (configurations). We formulate
the solution (configuration) counting problem within the framework
of inverse Ising problem, and this idea is tested on both diluted
models and fully-connected models, as well as on real neural data.
In the first case, we do not know a priori the underlying graphical
models and try to construct a disordered Ising model from the
collected data to evaluate the entropy of those unknown models. Note
that the sampled solutions come from the zero energy ground state
(single solution cluster) and the number is limited to $P$ but the
size of that sampled cluster is evaluated. To this end, the pairwise
model improves substantially the independent model (see
Fig.~\ref{compa}(a)). When the constraint density is small then the
pairwise correlation dominates the solution space, the estimated
entropy gets very close to the true one estimated with the knowledge
of the original model. Another interesting point to be demonstrated
in our further work is, for small $N$, one can compute the
magnetizations and correlations through exact enumeration and
further the real entropy value. The result of Boltzmann learning
algorithm~\cite{Ack-1985} and Monte Carlo
simulation~\cite{Bialek-09ep} (using a large enough amount of Monte
Carlo samplings) should provide an upper bound on the true entropy.
Instead, using the approximate method we proposed, whether the
obtained result provides a bound should be checked for small size
system or proved in the limit of large $N$, provided that the
magnetizations and correlations are less noisy. In the second case,
the susceptibility propagation is applied to infer the retinal
network and belief propagation is used to reproduce the entropy
computed by Monte Carlo method. This message passing scheme is very
fast and efficient especially for large network and the observed
multiple fixed points predict other metastable states in the
inferred retinal network. These metastable states may have intimate
relation with the neuronal population coding~\cite{Tkacik-2010}.
Extensions to the neuronal interaction network organized in a
hierarchical and modular manner would be very
interesting~\cite{Nature-10,Ganmor-11}. Our presented framework
constructs a statistical mechanics description of the system
directly from either artificial data or real data, and has the
potential to describe biological networks more generally and
estimate the size of the solution space in various contexts
especially when pairwise correlation dominates the system.
%%%%%%%%%%%%%%%%%%%%%%%%%%%%%%%%%%%%%%%%%%%%%%%%%%%%%%%
\section*{Acknowledgments}

%\begin{acknowledgments}
We thank Gasper Tkacik for providing us multielectrode recordings of
the salamander retina. The improvement of the manuscript benefited
from comments and suggestions of anonymous referees. The present
work was partially supported by the NSFC Grant 10834014 and the
973-Program Grant 2007CB935903 and HKUST 605010.
%\end{acknowledgments}
%%%%%%%%%%%%%%%%%%%%%%%%%%%%%%%%%%%%%%%%%%%%%%%%%%%%%%%%%%%%%%%
%\bibliography{ref}

\begin{thebibliography}{36}
\expandafter\ifx\csname
natexlab\endcsname\relax\def\natexlab#1{#1}\fi
\expandafter\ifx\csname bibnamefont\endcsname\relax
  \def\bibnamefont#1{#1}\fi
\expandafter\ifx\csname bibfnamefont\endcsname\relax
  \def\bibfnamefont#1{#1}\fi
\expandafter\ifx\csname citenamefont\endcsname\relax
  \def\citenamefont#1{#1}\fi
\expandafter\ifx\csname url\endcsname\relax
  \def\url#1{\texttt{#1}}\fi
\expandafter\ifx\csname urlprefix\endcsname\relax\def\urlprefix{URL
}\fi \providecommand{\bibinfo}[2]{#2}
\providecommand{\eprint}[2][]{\url{#2}}

\bibitem[{\citenamefont{Valiant}(1979)}]{Valiant-1979}
\bibinfo{author}{\bibfnamefont{L.}~\bibnamefont{Valiant}},
  \bibinfo{journal}{Theoretical Computer Sciences}
  \textbf{\bibinfo{volume}{8}}, \bibinfo{pages}{189} (\bibinfo{year}{1979}).

\bibitem[{\citenamefont{Roth}(1996)}]{Roth-1996}
\bibinfo{author}{\bibfnamefont{D.}~\bibnamefont{Roth}},
  \bibinfo{journal}{Artificial Intelligence} \textbf{\bibinfo{volume}{82}},
  \bibinfo{pages}{273} (\bibinfo{year}{1996}).

\bibitem[{\citenamefont{Gomes et~al.}(2007)\citenamefont{Gomes, Hoffmann,
  Sabharwal, and Selman}}]{Selman-2007}
\bibinfo{author}{\bibfnamefont{C.~P.} \bibnamefont{Gomes}},
  \bibinfo{author}{\bibfnamefont{J.}~\bibnamefont{Hoffmann}},
  \bibinfo{author}{\bibfnamefont{A.}~\bibnamefont{Sabharwal}},
  \bibnamefont{and} \bibinfo{author}{\bibfnamefont{B.}~\bibnamefont{Selman}},
  in \emph{\bibinfo{booktitle}{Proc. of IJCAI-07}}
  (\bibinfo{address}{Hyderabad, India}, \bibinfo{year}{2007}), pp.
  \bibinfo{pages}{2293--2299}.

\bibitem[{\citenamefont{Kroc et~al.}(2008)\citenamefont{Kroc, Sabharwal, and
  Selman}}]{Kroc-2008}
\bibinfo{author}{\bibfnamefont{L.}~\bibnamefont{Kroc}},
  \bibinfo{author}{\bibfnamefont{A.}~\bibnamefont{Sabharwal}},
  \bibnamefont{and} \bibinfo{author}{\bibfnamefont{B.}~\bibnamefont{Selman}},
  in \emph{\bibinfo{booktitle}{Proc. of CPAIOR-08}} (\bibinfo{address}{Paris,
  France}, \bibinfo{year}{2008}), pp. \bibinfo{pages}{127--141}.

\bibitem[{\citenamefont{Favier et~al.}(2009)\citenamefont{Favier, de~Givry, and
  J\'egou}}]{Favier-2009}
\bibinfo{author}{\bibfnamefont{A.}~\bibnamefont{Favier}},
  \bibinfo{author}{\bibfnamefont{S.}~\bibnamefont{de~Givry}}, \bibnamefont{and}
  \bibinfo{author}{\bibfnamefont{P.}~\bibnamefont{J\'egou}}, in
  \emph{\bibinfo{booktitle}{Proc. of CP-09}} (\bibinfo{address}{Lisbon,
  Portugal}, \bibinfo{year}{2009}), pp. \bibinfo{pages}{335--343}.

\bibitem[{\citenamefont{Bento and Montanari}(2009)}]{Montanari-2009}
\bibinfo{author}{\bibfnamefont{J.}~\bibnamefont{Bento}} \bibnamefont{and}
  \bibinfo{author}{\bibfnamefont{A.}~\bibnamefont{Montanari}}, in
  \emph{\bibinfo{booktitle}{Advances in Neural Information Processing Systems
  22}}, edited by \bibinfo{editor}{\bibfnamefont{Y.}~\bibnamefont{Bengio}},
  \bibinfo{editor}{\bibfnamefont{D.}~\bibnamefont{Schuurmans}},
  \bibinfo{editor}{\bibfnamefont{J.}~\bibnamefont{Lafferty}},
  \bibinfo{editor}{\bibfnamefont{C.~K.~I.} \bibnamefont{Williams}},
  \bibnamefont{and} \bibinfo{editor}{\bibfnamefont{A.}~\bibnamefont{Culotta}}
  (\bibinfo{year}{2009}), pp. \bibinfo{pages}{1303--1311}.

\bibitem[{\citenamefont{Schneidman et~al.}(2006)\citenamefont{Schneidman,
  Berry, Segev, and Bialek}}]{Nature-06}
\bibinfo{author}{\bibfnamefont{E.}~\bibnamefont{Schneidman}},
  \bibinfo{author}{\bibfnamefont{M.~J.} \bibnamefont{Berry}},
  \bibinfo{author}{\bibfnamefont{R.}~\bibnamefont{Segev}}, \bibnamefont{and}
  \bibinfo{author}{\bibfnamefont{W.}~\bibnamefont{Bialek}},
  \bibinfo{journal}{Nature} \textbf{\bibinfo{volume}{440}},
  \bibinfo{pages}{1007} (\bibinfo{year}{2006}).

\bibitem[{\citenamefont{Tkacik et~al.}(2009)\citenamefont{Tkacik, Schneidman,
  Berry, and Bialek}}]{Bialek-09ep}
\bibinfo{author}{\bibfnamefont{G.}~\bibnamefont{Tkacik}},
  \bibinfo{author}{\bibfnamefont{E.}~\bibnamefont{Schneidman}},
  \bibinfo{author}{\bibfnamefont{M.~J.} \bibnamefont{Berry}}, \bibnamefont{and}
  \bibinfo{author}{\bibfnamefont{W.}~\bibnamefont{Bialek}}
  (\bibinfo{year}{2009}), \bibinfo{note}{e-print arXiv:0912.5409}.

\bibitem[{\citenamefont{Cocco and Monasson}(2011)}]{Cocco-2011}
\bibinfo{author}{\bibfnamefont{S.}~\bibnamefont{Cocco}} \bibnamefont{and}
  \bibinfo{author}{\bibfnamefont{R.}~\bibnamefont{Monasson}},
  \bibinfo{journal}{Phys. Rev. Lett} \textbf{\bibinfo{volume}{106}},
  \bibinfo{pages}{090601} (\bibinfo{year}{2011}).

\bibitem[{\citenamefont{Macke et~al.}(2011)\citenamefont{Macke, Opper, and
  Bethge}}]{Opper-2011}
\bibinfo{author}{\bibfnamefont{J.~H.} \bibnamefont{Macke}},
  \bibinfo{author}{\bibfnamefont{M.}~\bibnamefont{Opper}}, \bibnamefont{and}
  \bibinfo{author}{\bibfnamefont{M.}~\bibnamefont{Bethge}},
  \bibinfo{journal}{Phys. Rev. Lett} \textbf{\bibinfo{volume}{106}},
  \bibinfo{pages}{208102} (\bibinfo{year}{2011}).

\bibitem[{\citenamefont{Ackley et~al.}(1985)\citenamefont{Ackley, Hinton, and
  Sejnowski}}]{Ack-1985}
\bibinfo{author}{\bibfnamefont{D.~H.} \bibnamefont{Ackley}},
  \bibinfo{author}{\bibfnamefont{G.~E.} \bibnamefont{Hinton}},
  \bibnamefont{and} \bibinfo{author}{\bibfnamefont{T.~J.}
  \bibnamefont{Sejnowski}}, \bibinfo{journal}{Cognitive Science}
  \textbf{\bibinfo{volume}{9}}, \bibinfo{pages}{147} (\bibinfo{year}{1985}).

\bibitem[{\citenamefont{M\'ezard and Mora}(2009)}]{Mezard-09}
\bibinfo{author}{\bibfnamefont{M.}~\bibnamefont{M\'ezard}} \bibnamefont{and}
  \bibinfo{author}{\bibfnamefont{T.}~\bibnamefont{Mora}}, \bibinfo{journal}{J.
  Physiology Paris} \textbf{\bibinfo{volume}{103}}, \bibinfo{pages}{107}
  (\bibinfo{year}{2009}).

\bibitem[{\citenamefont{Roudi et~al.}(2009)\citenamefont{Roudi, Tyrcha, and
  Hertz}}]{Roudi-2009}
\bibinfo{author}{\bibfnamefont{Y.}~\bibnamefont{Roudi}},
  \bibinfo{author}{\bibfnamefont{J.}~\bibnamefont{Tyrcha}}, \bibnamefont{and}
  \bibinfo{author}{\bibfnamefont{J.}~\bibnamefont{Hertz}},
  \bibinfo{journal}{Phys. Rev. E} \textbf{\bibinfo{volume}{79}},
  \bibinfo{pages}{051915} (\bibinfo{year}{2009}).

\bibitem[{\citenamefont{Sessak and Monasson}(2009)}]{SM-09}
\bibinfo{author}{\bibfnamefont{V.}~\bibnamefont{Sessak}} \bibnamefont{and}
  \bibinfo{author}{\bibfnamefont{R.}~\bibnamefont{Monasson}},
  \bibinfo{journal}{J. Phys. A} \textbf{\bibinfo{volume}{42}},
  \bibinfo{pages}{055001} (\bibinfo{year}{2009}).

\bibitem[{\citenamefont{Weigt et~al.}(2009)\citenamefont{Weigt, White,
  Szurmant, Hoch, and Hwa}}]{Weigt-2009}
\bibinfo{author}{\bibfnamefont{M.}~\bibnamefont{Weigt}},
  \bibinfo{author}{\bibfnamefont{R.~A.} \bibnamefont{White}},
  \bibinfo{author}{\bibfnamefont{H.}~\bibnamefont{Szurmant}},
  \bibinfo{author}{\bibfnamefont{J.~A.} \bibnamefont{Hoch}}, \bibnamefont{and}
  \bibinfo{author}{\bibfnamefont{T.}~\bibnamefont{Hwa}},
  \bibinfo{journal}{Proc. Natl. Acad. Sci. USA} \textbf{\bibinfo{volume}{106}},
  \bibinfo{pages}{67} (\bibinfo{year}{2009}).

\bibitem[{\citenamefont{Cocco et~al.}(2009)\citenamefont{Cocco, Leibler, and
  Monasson}}]{Cocco-09}
\bibinfo{author}{\bibfnamefont{S.}~\bibnamefont{Cocco}},
  \bibinfo{author}{\bibfnamefont{S.}~\bibnamefont{Leibler}}, \bibnamefont{and}
  \bibinfo{author}{\bibfnamefont{R.}~\bibnamefont{Monasson}},
  \bibinfo{journal}{Proc. Natl. Acad. Sci. USA} \textbf{\bibinfo{volume}{106}},
  \bibinfo{pages}{14058} (\bibinfo{year}{2009}).

\bibitem[{\citenamefont{Mora et~al.}(2010)\citenamefont{Mora, Walczak, Bialek,
  and C.~G.~Callan}}]{Mora-2010}
\bibinfo{author}{\bibfnamefont{T.}~\bibnamefont{Mora}},
  \bibinfo{author}{\bibfnamefont{A.~M.} \bibnamefont{Walczak}},
  \bibinfo{author}{\bibfnamefont{W.}~\bibnamefont{Bialek}}, \bibnamefont{and}
  \bibinfo{author}{\bibfnamefont{J.}~\bibnamefont{C.~G.~Callan}},
  \bibinfo{journal}{Proc. Natl. Acad. Sci. USA} \textbf{\bibinfo{volume}{107}},
  \bibinfo{pages}{5405} (\bibinfo{year}{2010}).

\bibitem[{\citenamefont{Tang et~al.}(2008)\citenamefont{Tang, Jackson, Hobbs,
  Chen, Smith, Patel, Prieto, Petrusca, Grivich, Sher et~al.}}]{Tang-2008}
\bibinfo{author}{\bibfnamefont{A.}~\bibnamefont{Tang}},
  \bibinfo{author}{\bibfnamefont{D.}~\bibnamefont{Jackson}},
  \bibinfo{author}{\bibfnamefont{J.}~\bibnamefont{Hobbs}},
  \bibinfo{author}{\bibfnamefont{W.}~\bibnamefont{Chen}},
  \bibinfo{author}{\bibfnamefont{J.~L.} \bibnamefont{Smith}},
  \bibinfo{author}{\bibfnamefont{H.}~\bibnamefont{Patel}},
  \bibinfo{author}{\bibfnamefont{A.}~\bibnamefont{Prieto}},
  \bibinfo{author}{\bibfnamefont{D.}~\bibnamefont{Petrusca}},
  \bibinfo{author}{\bibfnamefont{M.~I.} \bibnamefont{Grivich}},
  \bibinfo{author}{\bibfnamefont{A.}~\bibnamefont{Sher}}, \bibnamefont{et~al.},
  \bibinfo{journal}{J. Neurosci} \textbf{\bibinfo{volume}{28}},
  \bibinfo{pages}{505} (\bibinfo{year}{2008}).

\bibitem[{\citenamefont{Tkacik et~al.}(2010)\citenamefont{Tkacik, Prentice,
  Balasubramanian, and Schneidman}}]{Tkacik-2010}
\bibinfo{author}{\bibfnamefont{G.}~\bibnamefont{Tkacik}},
  \bibinfo{author}{\bibfnamefont{J.~S.} \bibnamefont{Prentice}},
  \bibinfo{author}{\bibfnamefont{V.}~\bibnamefont{Balasubramanian}},
  \bibnamefont{and}
  \bibinfo{author}{\bibfnamefont{E.}~\bibnamefont{Schneidman}},
  \bibinfo{journal}{Proc. Natl. Acad. Sci. USA} \textbf{\bibinfo{volume}{107}},
  \bibinfo{pages}{14419} (\bibinfo{year}{2010}).

\bibitem[{\citenamefont{Marinari and Kerrebroeck}(2010)}]{Marinari-2010}
\bibinfo{author}{\bibfnamefont{E.}~\bibnamefont{Marinari}} \bibnamefont{and}
  \bibinfo{author}{\bibfnamefont{V.~V.} \bibnamefont{Kerrebroeck}},
  \bibinfo{journal}{J. Stat. Mech.: Theory Exp}
  \textbf{\bibinfo{volume}{P02008}} (\bibinfo{year}{2010}).

\bibitem[{\citenamefont{Huang}(2010)}]{Huang-2010b}
\bibinfo{author}{\bibfnamefont{H.}~\bibnamefont{Huang}},
  \bibinfo{journal}{Phys. Rev. E} \textbf{\bibinfo{volume}{82}},
  \bibinfo{pages}{056111} (\bibinfo{year}{2010}).

\bibitem[{\citenamefont{M\'ezard and Parisi}(2001)}]{cavity-2001}
\bibinfo{author}{\bibfnamefont{M.}~\bibnamefont{M\'ezard}} \bibnamefont{and}
  \bibinfo{author}{\bibfnamefont{G.}~\bibnamefont{Parisi}},
  \bibinfo{journal}{Eur. Phys. J. B} \textbf{\bibinfo{volume}{20}},
  \bibinfo{pages}{217} (\bibinfo{year}{2001}).

\bibitem[{\citenamefont{Li et~al.}(2009)\citenamefont{Li, Ma, and
  Zhou}}]{Zhou-2009}
\bibinfo{author}{\bibfnamefont{K.}~\bibnamefont{Li}},
  \bibinfo{author}{\bibfnamefont{H.}~\bibnamefont{Ma}}, \bibnamefont{and}
  \bibinfo{author}{\bibfnamefont{H.}~\bibnamefont{Zhou}},
  \bibinfo{journal}{Phys. Rev. E} \textbf{\bibinfo{volume}{79}},
  \bibinfo{pages}{031102} (\bibinfo{year}{2009}).

\bibitem[{\citenamefont{Braunstein and Zecchina}(2006)}]{Zecchina-2006}
\bibinfo{author}{\bibfnamefont{A.}~\bibnamefont{Braunstein}} \bibnamefont{and}
  \bibinfo{author}{\bibfnamefont{R.}~\bibnamefont{Zecchina}},
  \bibinfo{journal}{Phys. Rev. Lett} \textbf{\bibinfo{volume}{96}},
  \bibinfo{pages}{030201} (\bibinfo{year}{2006}).

\bibitem[{\citenamefont{Monasson and Zecchina}(1996)}]{Monasson-1996}
\bibinfo{author}{\bibfnamefont{R.}~\bibnamefont{Monasson}} \bibnamefont{and}
  \bibinfo{author}{\bibfnamefont{R.}~\bibnamefont{Zecchina}},
  \bibinfo{journal}{Phys. Rev. Lett} \textbf{\bibinfo{volume}{76}},
  \bibinfo{pages}{3881} (\bibinfo{year}{1996}).

\bibitem[{\citenamefont{Krzakala et~al.}(2007)\citenamefont{Krzakala,
  Montanari, {Ricci-Tersenghi}, Semerjian, and Zdeborova}}]{Krzakala-PNAS-2007}
\bibinfo{author}{\bibfnamefont{F.}~\bibnamefont{Krzakala}},
  \bibinfo{author}{\bibfnamefont{A.}~\bibnamefont{Montanari}},
  \bibinfo{author}{\bibfnamefont{F.}~\bibnamefont{{Ricci-Tersenghi}}},
  \bibinfo{author}{\bibfnamefont{G.}~\bibnamefont{Semerjian}},
  \bibnamefont{and}
  \bibinfo{author}{\bibfnamefont{L.}~\bibnamefont{Zdeborova}},
  \bibinfo{journal}{Proc. Natl. Acad. Sci. USA} \textbf{\bibinfo{volume}{104}},
  \bibinfo{pages}{10318} (\bibinfo{year}{2007}).

\bibitem[{\citenamefont{Zhou}(2010)}]{Zhou-2010epjb}
\bibinfo{author}{\bibfnamefont{H.}~\bibnamefont{Zhou}}, \bibinfo{journal}{Eur.
  Phys. J. B} \textbf{\bibinfo{volume}{73}}, \bibinfo{pages}{617}
  (\bibinfo{year}{2010}).

\bibitem[{\citenamefont{Krauth and M\'ezard}(1989)}]{Krauth-1989}
\bibinfo{author}{\bibfnamefont{W.}~\bibnamefont{Krauth}} \bibnamefont{and}
  \bibinfo{author}{\bibfnamefont{M.}~\bibnamefont{M\'ezard}},
  \bibinfo{journal}{J. Phys. (France)} \textbf{\bibinfo{volume}{50}},
  \bibinfo{pages}{3057} (\bibinfo{year}{1989}).

\bibitem[{\citenamefont{Huang and Zhou}(2010)}]{Huang-2010jstat}
\bibinfo{author}{\bibfnamefont{H.}~\bibnamefont{Huang}} \bibnamefont{and}
  \bibinfo{author}{\bibfnamefont{H.}~\bibnamefont{Zhou}}, \bibinfo{journal}{J.
  Stat. Mech.: Theory Exp} \textbf{\bibinfo{volume}{P08014}}
  (\bibinfo{year}{2010}).

\bibitem[{\citenamefont{Schneidman et~al.}(2003)\citenamefont{Schneidman,
  Still, Berry, and Bialek}}]{Bialek-03}
\bibinfo{author}{\bibfnamefont{E.}~\bibnamefont{Schneidman}},
  \bibinfo{author}{\bibfnamefont{S.}~\bibnamefont{Still}},
  \bibinfo{author}{\bibfnamefont{M.~J.} \bibnamefont{Berry}}, \bibnamefont{and}
  \bibinfo{author}{\bibfnamefont{W.}~\bibnamefont{Bialek}},
  \bibinfo{journal}{Phys. Rev. Lett} \textbf{\bibinfo{volume}{91}},
  \bibinfo{pages}{238701} (\bibinfo{year}{2003}).

\bibitem[{\citenamefont{Higuchi and M\'ezard}(2010)}]{Higuchi-2010}
\bibinfo{author}{\bibfnamefont{S.}~\bibnamefont{Higuchi}} \bibnamefont{and}
  \bibinfo{author}{\bibfnamefont{M.}~\bibnamefont{M\'ezard}},
  \bibinfo{journal}{J. Phys.: Conf. Ser} \textbf{\bibinfo{volume}{233}},
  \bibinfo{pages}{012003} (\bibinfo{year}{2010}).

\bibitem[{\citenamefont{Zhou and Ma}(2009)}]{Zhou-2009pre}
\bibinfo{author}{\bibfnamefont{H.}~\bibnamefont{Zhou}} \bibnamefont{and}
  \bibinfo{author}{\bibfnamefont{H.}~\bibnamefont{Ma}}, \bibinfo{journal}{Phys.
  Rev. E} \textbf{\bibinfo{volume}{80}}, \bibinfo{pages}{066108}
  (\bibinfo{year}{2009}).

\bibitem[{\citenamefont{Mann and Hartmann}(2010)}]{Hartmann-2010}
\bibinfo{author}{\bibfnamefont{A.}~\bibnamefont{Mann}} \bibnamefont{and}
  \bibinfo{author}{\bibfnamefont{A.~K.} \bibnamefont{Hartmann}},
  \bibinfo{journal}{Phys. Rev. E} \textbf{\bibinfo{volume}{82}},
  \bibinfo{pages}{056702} (\bibinfo{year}{2010}).

\bibitem[{\citenamefont{Huang and Zhou}(2011)}]{Huang-2011epl}
\bibinfo{author}{\bibfnamefont{H.}~\bibnamefont{Huang}} \bibnamefont{and}
  \bibinfo{author}{\bibfnamefont{H.}~\bibnamefont{Zhou}},
  \bibinfo{journal}{Europhys. Lett} \textbf{\bibinfo{volume}{96}},
  \bibinfo{pages}{58003} (\bibinfo{year}{2011}).

\bibitem[{\citenamefont{Ohiorhenuan et~al.}(2010)\citenamefont{Ohiorhenuan,
  Mechler, Purpura, Schmid, Hu, and Victor}}]{Nature-10}
\bibinfo{author}{\bibfnamefont{I.~E.} \bibnamefont{Ohiorhenuan}},
  \bibinfo{author}{\bibfnamefont{F.}~\bibnamefont{Mechler}},
  \bibinfo{author}{\bibfnamefont{K.~P.} \bibnamefont{Purpura}},
  \bibinfo{author}{\bibfnamefont{A.~M.} \bibnamefont{Schmid}},
  \bibinfo{author}{\bibfnamefont{Q.}~\bibnamefont{Hu}}, \bibnamefont{and}
  \bibinfo{author}{\bibfnamefont{J.~D.} \bibnamefont{Victor}},
  \bibinfo{journal}{Nature} \textbf{\bibinfo{volume}{466}},
  \bibinfo{pages}{617} (\bibinfo{year}{2010}).

\bibitem[{\citenamefont{Ganmor et~al.}(2011)\citenamefont{Ganmor, Segev, and
  Schneidman}}]{Ganmor-11}
\bibinfo{author}{\bibfnamefont{E.}~\bibnamefont{Ganmor}},
  \bibinfo{author}{\bibfnamefont{R.}~\bibnamefont{Segev}}, \bibnamefont{and}
  \bibinfo{author}{\bibfnamefont{E.}~\bibnamefont{Schneidman}},
  \bibinfo{journal}{Proc. Natl. Acad. Sci. USA} \textbf{\bibinfo{volume}{108}},
  \bibinfo{pages}{9679} (\bibinfo{year}{2011}).

\end{thebibliography}

%%%%%%%%%%%%%%%%%%%%%%%%%%%%%%%%%%%%%%%%%%%%%%%%%%%%%%%%%%%%%%%%%%%%%

\end{document}